\documentclass{jfm}
\usepackage{graphicx}
\usepackage{epstopdf, epsfig}

\shorttitle{Near-surface coherent structures in an intense tropical cyclone}
\shortauthor{C. N. Oguejiofor and D. H. Richter}

\title{Near-surface coherent structures in an intense tropical cyclone: conditional eddies and vertical momentum fluxes}

\author{Chibueze N. Oguejiofor\aff{1}
  \corresp{\email{coguejio@nd.edu}}
 \and David H. Richter\aff{1}
 }

\affiliation{\aff{1}University of Notre Dame, South Bend, Indiana
}

\begin{document}

\maketitle


\begin{abstract}
The intermittency of coherent turbulent structures in the tropical cyclone boundary layer makes them challenging to fully characterize, especially regarding their impact on momentum dynamics in the eyewall. Furthermore, the fine spatial and temporal model resolution needed to resolve these structures has long impeded their understanding. Using the output of a large eddy simulation (with 31-m horizontal and 15-m vertical grid spacing), we investigate the three-dimensional structure of conditionally averaged $eddies$ associated with extreme Reynolds stress events in the near-surface region of a category 5 tropical cyclone. For ejection events, we find (using two- and three-dimensional analysis) that the structure of the conditional eddy is a vortex core roughly inclined toward the direction of the mean flow in the inflowing boundary layer, outer and inner eyewall region. For sweep events, this vortex core is less coherent. The nature of the educed eddies found in this study suggests a similarity between the coherent structures in simpler wall-bounded turbulent flows and those in complex environmental flows.
\end{abstract}


%
\section{Introduction}\label{sec:sect1_intro}
Traditional approaches to turbulence research in meteorology, involving Reynolds averaging and other eddy-mean partitioning methodology, only ever address turbulence in the \textit{mean} sense; rarely confronting the intermittency of coherent turbulent structures and their potential roles. We ask: in the context of hurricane dynamics, how much of our current understanding of the role of turbulence remains constrained by this bulk treatment of turbulent processes? This question becomes increasingly pertinent in the face of various observational evidence of organized turbulent structures in the hurricane boundary layer (HBL).

The first observational evidence of coherent structures in the HBL was reported by \cite{wurman1998intense} in their study of Hurricane Fran (1996).
Using the high resolution Doppler on wheels (DOW) ($\approx$75m resolution), they noted the presence of sub-kilometer scale roll vortices, approximately aligned with the mean tangential winds. In the years since that paper was published, the improvement of remote sensing technologies enabled the identification of HBL roll vortices in ground-based doppler radars \citep{morrison2005observational,lorsolo2008observational}, spaceborne sythentic aperture radar (SAR) imagery \citep{li2013tropical,horstmann2015tropical,huang2018tropical,combot2020extensive} and aircrafts \citep{zhang2008effects,tang2021direct}. A rigorous examination of such structures and a theoretical framework for their prediction was presented by \cite{foster2005rolls}, using linear stability theory. Several modeling studies \citep{nakanishi2012large,gao2014generation,gao2016equilibrium,gao2017effect,gao2018characteristics,li2021effects,li2023dynamic} have since been carried out in an attempt to understand the origin and influence of roll vortices in the HBL, with a growing consensus on their role in enhancing momentum fluxes. Although roll vortices appear to be the most discussed form of coherent structure in the HBL, several other types have also been documented e.g., inner-core fingers/striations \citep{bluestein1987structure,aberson2006hurricane,tsukada2020estimation}, mesovortices \citep{kossin2002vortical,kossin2004mesovortices,wurman2018role,alford2023using}, tornado scale vortices \citep{wurman2018role}, coherent turbulent eddies \citep{guimond2018coherent,protzko2023documenting} etc.

The study of coherent structures in the near-wall turbulent region, however dates farther back to it's first mention by \cite{theodorsen1952mechanisms,theodorsen1955structure} when \textit{horse-shoe}--like vortex filament structures were investigated using smoke visualizations and shown to participate actively in fluid transport. 
Reports of streaks (zones of comparatively weak streamwise momentum) in the near wall region were then presented by \cite{kline1967structure} using their Hydrogen bubbles experiments.
Thereafter, \cite{townsend1976structure} proposed the \textit{attached eddy} phenomenology for dominant structures in the turbulent near-wall region, defining "active motions" as ones leading to strong correlations in turbulent velocity components, producing Reynolds stress.
Using large eddy simulation (LES), a series of studies by \cite{moin1982numerical,moin1985structure,kim1986structure} showed, for the first time, the existence of hairpin vortices and omega ($\Omega$) like structures in instantaneous fields of a turbulent channel flow. The direct numerical simulation (DNS)  of a low Reynolds number boundary layer \citep{spalart1988direct} was extensively analyzed by \cite{robinson1991kinematics,robinson1991coherent}, thereafter noting the presence of a suite of structures: quasi-streamwise vortices (rolls) and archs with the horse-shoe/hairpin vortices being less frequently seen as previously suggested. Several studies have since then explored the existence of coherent structures in a variety of simple wall-bounded engineering turbulent flows, emphasizing the roles they play in enhancing momentum transport \citep{hussain1981role,bernard1990reynolds,smith1991dynamics,handler1992role,bernard1993vortex,lozano2012three,encinar2020momentum,guerrero2020extreme,shah2021three,wan2023conditional}. However, these evidences for the prevalence of coherent structures in simple flows do not imply the existence of similar structures in more complex, realistic environmental flows.  

Using a LES, \cite{lin1996coherent} investigated the dynamics of coherent structures in a neutrally stratified planetary boundary layer (PBL), documenting the existence of coherent vortical structures (horse-shoe \& quasi-streamwise) and their relationship to intense momentum fluxes. Along the same lines, \cite{foster2006near} showed that the coherent structures associated with extreme vertical momentum fluxes were closely related to the streaks seen in their 3D LES of a neutrally stratified atmospheric boundary layer (ABL). Recently, \cite{salesky2020coherent} showed that the inclusion of a velocity scale for Large and Very Large Scale Motions (LSMs and VLSMs) in a revised formulation of the Monin-Obukhov similarity theory (MOST) led to an improvement in the prediction of the atmospheric surface layer flux-gradient relationship, even though it had been previously noted by \cite{shih1987second} that models without explicitly resolved coherent structures accurately predict the mean Reynolds stress profiles. In much more complex flow regimes where coherent structures produced by the mean flow can be advected over considerable distances (e.g., in hurricanes -- Fig. \ref{fig:Fig0}), the prediction of momentum flux transport may be inaccurate without the explicit inclusion of the role of coherent structures. A similar idea was also alluded to by \cite{holmes2012turbulence} pg. 31.

The current study, thus, aims at understanding the relationship between coherent structures and intense vertical momentum flux occurrence in the HBL. To the authors' knowledge, this is the first time that a  treatment of the kinematics of coherent structures -- and their relationship to Reynolds Stress occurrence in different regions of the HBL -- have been presented in a turbulence resolving simulation of a hurricane. It is the hope of the authors' that this study would rekindle interests in the possible inclusion of the role of coherent structures in revised formulations of near-surface flux-gradient relationships in complex terrains. To this end, we pose the following research question: What form do the coherent structures responsible for extreme Reynolds stress events take, and how do they vary with storm location?

\section{Methodology}\label{sec:sect2_datamthd}
\subsection{Numerical Model And Dataset Description}\label{sec:sect2_datamthd_01}
This study uses a dataset from the LES of an idealized Category 5 hurricane detailed in \cite{worsnop2017using}, \cite{stern2018using}, \cite{stern2021estimating}, \cite{richter2021potential} and \cite{oguejiofor2024role}. The Cloud Model 1 (CM1) \citep{bryan2002benchmark,bryan2009maximum} was used with horizontal and vertical grid spacing of $\Delta x = \Delta y  $ = 31.25m, \  $\Delta z$ = 15.625m respectively, utilizing the eddy injection methodology described in detail in \cite{bryan2017eddy}.

\begin{figure}
    \centering
        \includegraphics[width=0.80\textwidth]{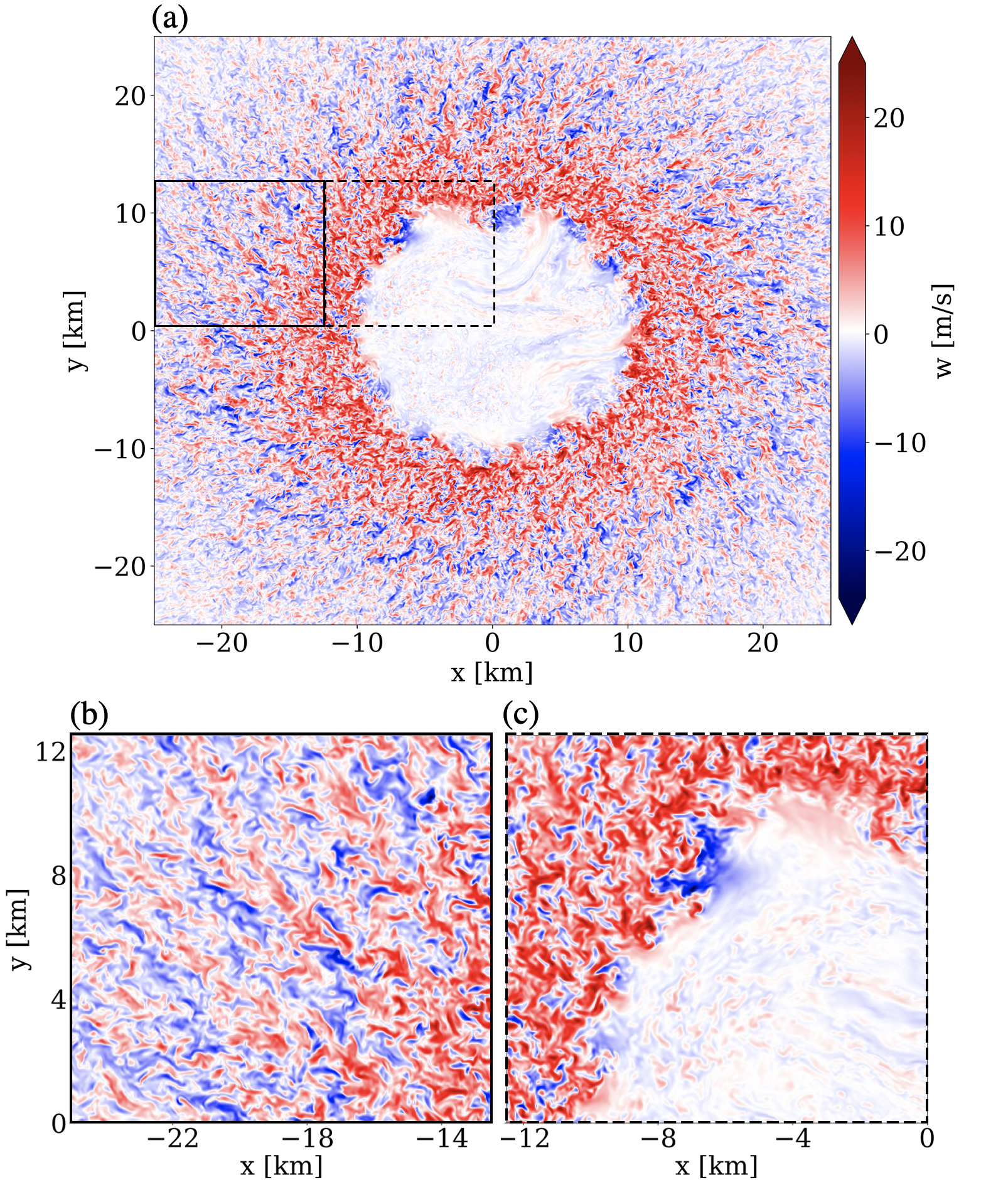}
    \caption{(a) Horizontal cross section of the vertical velocity field at z$\approx$700m. (b) and (c) show windowed-in regions of the N-W quadrant illustrating the organization of turbulent velocity structures in the outer eyewall and eye-eyewall interface respectively. [A plane view of the 3D vertical velocity isosurface (for (c)) is shown in Fig. 2 below (bottom--center panel i.e., ${T}_4$)].}
    \label{fig:Fig0}
\end{figure}

\begin{figure}
    \centering
        \includegraphics[width=0.99\textwidth]{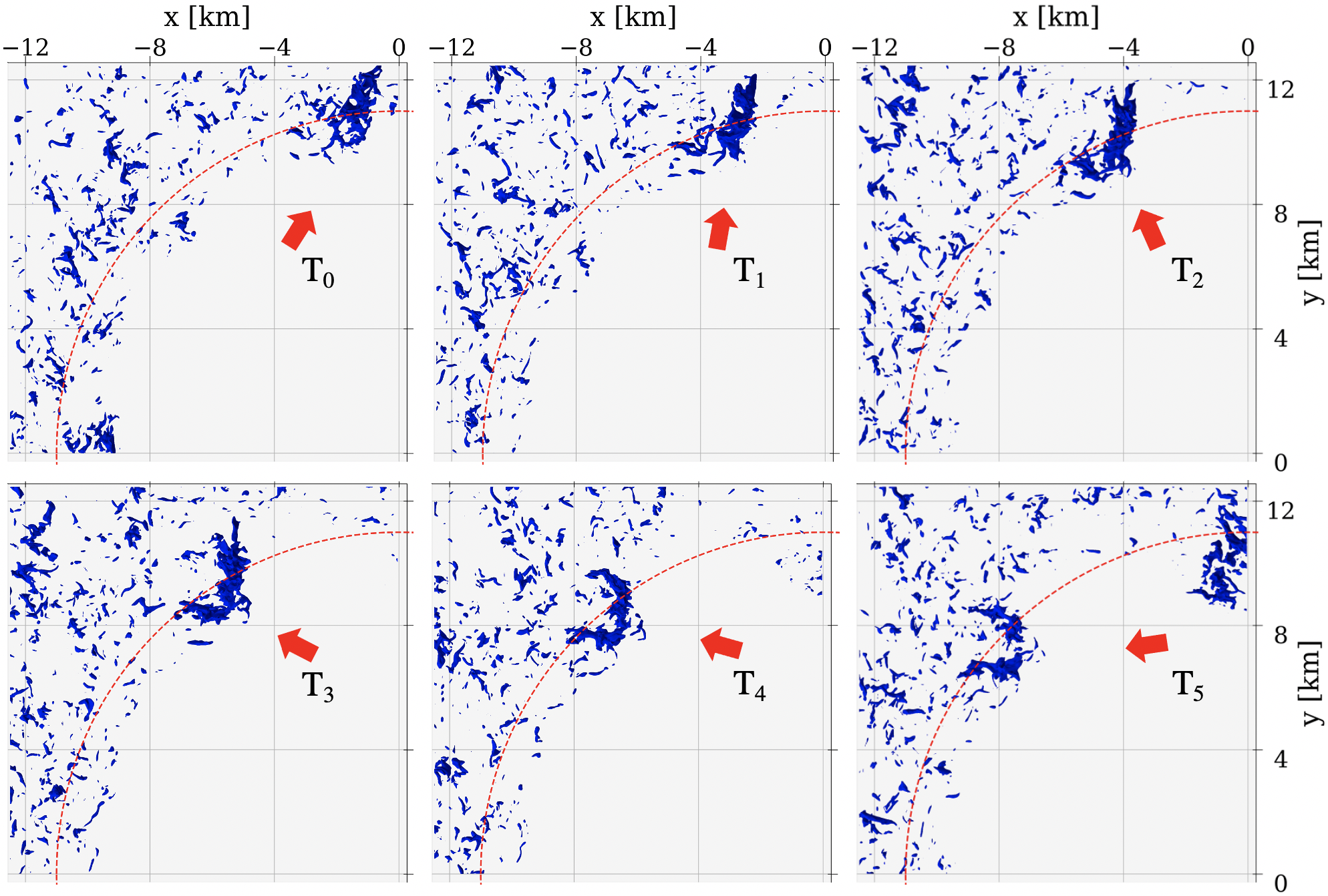}
    \caption{Three-dimensional isosurfaces (plan view) of instantaneous velocity structures (i.e., intense downdrafts with $w$=-8$ms^{-1}$) advected -- from ${T}_0$ to ${T}_5$ -- along the eye-eyewall interface of the simulated Category 5 Hurricane. The interval between each time step ($\Delta {T}$) is 20s and the red-dashed line in all panels marks r$\approx$11km, indicating the eye-eyewall interface. [The bottom-center panel, marked ${T}_4$, shows the same time instance as Fig. 1(c) above]. }
    \label{fig:Fig1}
\end{figure}

We focus on a 1hr evolution of the storm at steady state intensity, within a $\approx$50km $\times$ 50km subset (Fig. \ref{fig:Fig0}(a)) of the full LES subdomain 80km $\times$ 80km (see Figure 1 of \cite{oguejiofor2024role}). We analyze 61 snapshots of simulation output (1 minute temporal frequency) for the discussions herein. Figure \ref{fig:Fig0} shows a horizontal section of the vertical velocity field $w$ [m/s] at $\approx$700m height, showing the variety of resolved turbulent structures in the model output. These structures are qualitatively similar to those documented observationally (see a more detailed discussion in \cite{oguejiofor2024role}). Figures \ref{fig:Fig0}(b)-(c) shows the windowed-in views of the radial variation in the organization of turbulent structures (in the vertical velocity field at $\approx 700m$ height) between the inflowing boundary layer (Fig. \ref{fig:Fig0}(b)) and the eye-eyewall region (Fig. \ref{fig:Fig0}(c)). We note that in Fig. \ref{fig:Fig0}(b), the streaky pattern of turbulence organization is visible and has been documented extensively in marine atmospheric boundary layer (MABL) \citep{wang2019characteristics,wang2020assessment,stopa2022automated}. 
In Fig. 1(c) however, the vertical velocity field is less linearly organized, possibly due to the increased tangential velocity (compared to the radial velocity) and vertical velocity in the near-eyewall region. Qualitatively speaking, there is a clear transition in the flow regime between Fig. \ref{fig:Fig0}(b) and Fig. \ref{fig:Fig0}(c). The isosurfaces of vertical velocity ($w$=-8$ms^{-1}$), in Fig. \ref{fig:Fig1}, shows the temporal evolution of one prominent coherent velocity structure (identified in Fig. \ref{fig:Fig0}(c)) over $\approx$1min duration. Fig. \ref{fig:Fig1} shows clearly that there is a continuous cycle of "merging" and "disintegration" of turbulent structures in the eyewall of the simulated Tropical Cylcone (TC), supporting arguments for the intermittency of coherent structures. These velocity structures are most certainly qualitatively intriguing, however, the present study will be focussed on near-surface Reynolds stresses, as they are more relevant to the description of turbulence in the HBL.

For the analyses presented in this study, all 61 snapshots of instantaneous fields ($\mbox{\Large$\chi$} (x,y,z,t)$) are first time averaged in the native cartesian coordinate, afterwhich the turbulent fields are obtain by subtracting the time-mean field from each instantaneous fields $\mbox{\Large$\chi$}^{'}$ = $\mbox{\Large$\chi$}$ - $\langle $\mbox{\Large$\chi$}$ \rangle$. These turbulent fields are then interpolated into a cylindrical coordinates with radial grid points of r$=\Delta x$, $2\Delta x$, $3\Delta x$ $\dots$ to r$\approx$30km. There are about 6800 data points at each radii in the azimuthal direction after interpolation, sufficiently sampling the space which serve as data points. These interpolated turbulent fields (in cylindrical coordinates) are then used for our analyses.

\subsection{Conditional Averaging and Compositing}\label{sec:sect2_datamthd_03}
The closure problem in the Reynolds Averaged Navier-Stokes (RANS) momentum equations (see Eqns. 1 -- 3 in \cite{oguejiofor2024role}) necessitates the approximation of the ensuing averaged correlation between the turbulent velocity components (i.e., $\langle u^\prime w^\prime \rangle$, $\langle v^\prime w^\prime \rangle$, $\langle u^\prime v^\prime \rangle$, $\langle u^\prime u^\prime \rangle$, $\langle v^\prime v^\prime \rangle$, $\langle w^\prime w^\prime \rangle$), unlike in LES/DNS where these terms are resolved. The divergences of these correlation terms have been shown to play a significant role in modulating the TC mean wind fields, as clearly shown in \cite{oguejiofor2024role}. Furthermore, seeing that the product of the average of this correlation terms (Reynolds stresses) and the mean velocity gradient is responsible for the turbulence production (in the turbulent kinetic energy budget equation), a study of the kinematics of extreme Reynolds stress occurrences could provide insights on the production of turbulence in the HBL. 

The idea that the nature of eddies which produce Reynolds stresses could be educed by analysing the average flow field around a point in the flow, was promoted by \cite{adrian2007hairpin}.
Thus, by conditionally averaging the field around extreme Reynolds stress events, one can extract a "conditional eddy" representative of the average field associated with the Reynolds stress occurrence. 
At this point, we would like to emphasize that the educed conditional eddy structure is strongly dependent on the condition upon which the averaging is performed \citep{adrian1996stochastic}. Although extreme Reynolds stress events is a common metric used in educing conditionally averaged eddy structures \citep{foster2006near,lozano2012three,dong2017coherent,lozano2020non,guerrero2020extreme,maeyama2023near}, several other metrics for conditional averaging have been used in the past including: pressure perturbations in vegetation canopy flows, $p_T$ \citep{fitzmaurice2004three,finnigan2009turbulence,bailey2013turbulence}, inter-scale energy transfer, $\Pi$ \citep{hartel1994subgrid,natrajan2006role,hong2012coherent,dong2020coherent,wang2021coherent} etc. Seeing that the mechanism for the production of extreme Reynolds stress events is the focus of this study, it is imperative that we clearly define a "trigger" value to identify these events. We pick this trigger value  -- as the top 99.9 percentile -- from the distribution of the Reynolds stresses at a given height and radii range of interest. 

The methodology for conditional averaging used in this study follows a two-step procedure:

(i) Identification of the extreme occurrences of vertical momentum fluxes ($u^\prime w^\prime$, $v^\prime w^\prime$): At a given height in the domain, values of instantaneous vertical momentum fluxes exceeding a given trigger value are identified. Upon identification, a 1km $\times$ 1km $\times$ 0.3km rectangular volume is cut out. This volume of instantaneous turbulent field is centered around the extreme event, unless the height of analysis is close to the surface, in which case the volume lies above the extreme event. Grid points within this isolated volume are excluded from subsequent searches for more extreme Reynolds stress events. At the end of this procedure, a number of 3D volumes of instantaneous fields centered around extreme Reynolds Stress events are obtained.

(ii) Compositing: The 3D volumes of instantaneous turbulent fields from the above step are ensemble averaged in a new coordinate r$^{\prime}$[m], $\phi^{\prime}$[m], z$^{\prime}$[m]. This ensemble averaged volume represents the conditional eddy, from which quantities like vorticity ($\omega$), vortex identification criteria (e.g., Q-criteria) etc., are computed to further characterize the structure of the eddy.

\section{Results}\label{sec:sect3_resultsndisc}
\subsection{Near-surface mean velocity fields \&  turbulent fluxes}\label{sec:sec3dddy}
Before addressing the presence of coherent turbulent structures, we aim, first, to understand the mean wind conditions around the height where our analyses is focused (z$\approx$63m). For the discussions in the rest of this study, we divide this near surface region of the HBL into three distinct zones i.e., inner eyewall (r$\approx$10 - 13km), outer eyewall (r$\approx$14 - 17km) and inflowing BL (r$\approx$22 - 30km), based on behaviour of the mean velocity fields discussed below and the inflow angle (not shown).

\begin{figure}
    \centering
        \includegraphics[width=0.96\textwidth]{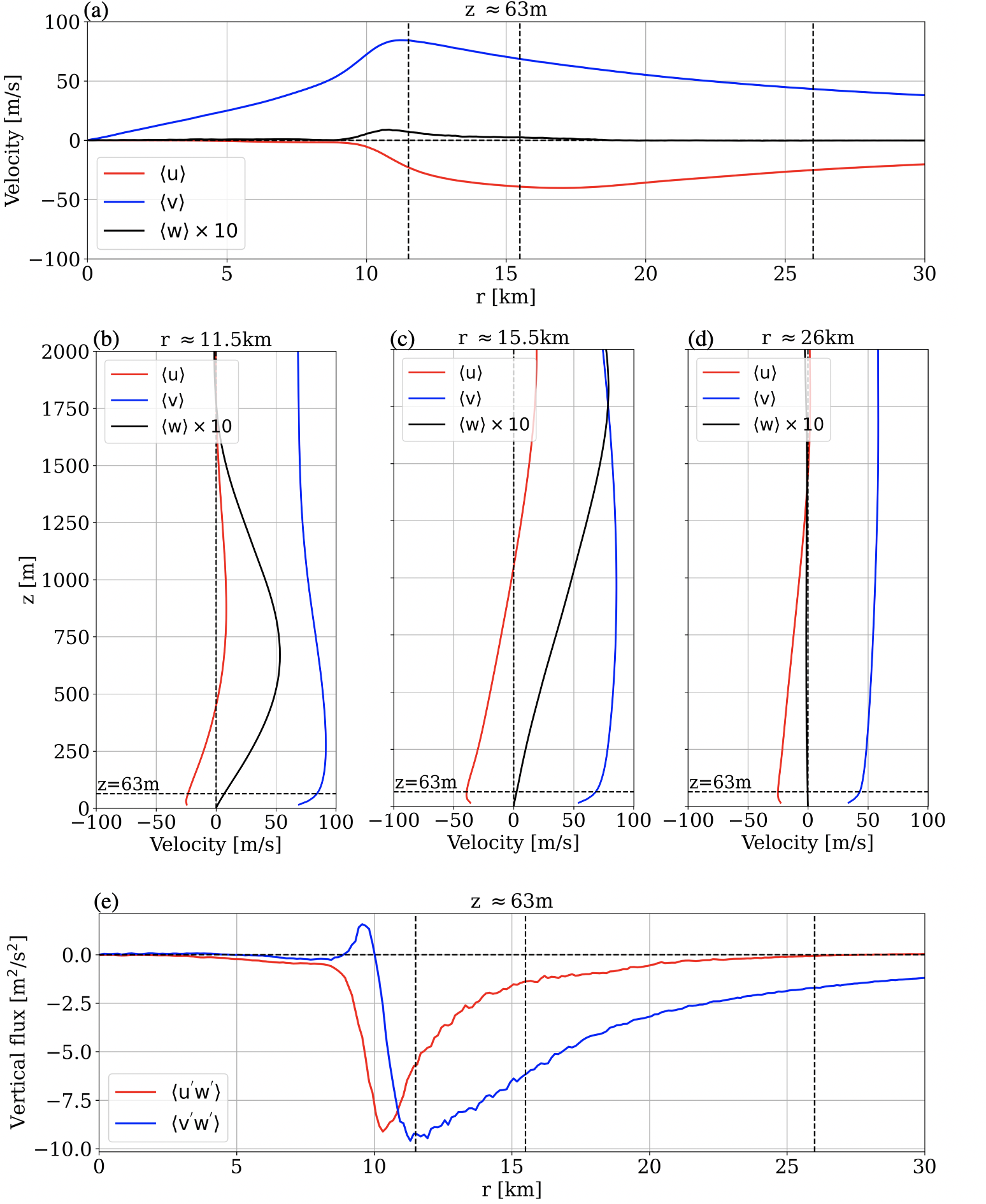}
    \caption{(a) Horizontal profiles of the azimuthally and time-averaged radial $\langle u \rangle$ (solid red line), tangential $\langle v \rangle$ (solid blue line) and vertical $\langle w \rangle$ [multiplied by a factor of 10 for visibility] (solid black line) velocity components (all in m/s) at z$\approx$63m height, with the dashed black lines indicating the radii (11.5km, 15.5km and 26km) where the vertical profiles in (b), (c) and (d) are plotted. (d) Horizontal profiles of the azimuthally and time-averaged vertical fluxes of radial velocity (red line -- $\langle u^\prime w^\prime \rangle$) and tangential velocity (blue line -- $\langle v^\prime w^\prime \rangle$).}
    \label{fig:Fig2}
\end{figure}

Figure \ref{fig:Fig2}(a) shows a horizontal profile of the mean velocity components. The radial velocity ($\langle u \rangle$ -- red solid line) is seen to be entirely negative, indicative of radial inflow towards the eye, peaking in outer the eyewall region, before decreasing towards the inner eyewall. The tangential velocity ($\langle v \rangle$ -- blue solid line) is maximum at the inner edge of the eyewall (r$\approx$11km), decreasing steadily away from the eye and sharply towards the eye. The vertical velocity ($\langle w \rangle$ -- black solid line) is seen to be near zero at this height, across all radii. For visibility, $\langle w \rangle$ plotted in Fig. \ref{fig:Fig2}(a) is multiplied by a factor of 10.0 and is seen to only become slightly significant towards the inner edge of the eyewall (i.e., r$\approx$11km). 
The analyses presented in this study focuses on understanding the near surface behaviour of coherent structures in regions of of HBL with notably different environmental conditions, partly to clarify how the kinematics of these structures may/may not vary across different flow regimes. Therefore three vertical profiles from the inner eyewall (Fig. \ref{fig:Fig2}(b)), outer eyewall (Fig. \ref{fig:Fig2}(c)) and inflowing BL (Fig. \ref{fig:Fig2}(d)) respectively are shown below. From these profiles, focussing on z$\approx$63m -- dashed black line, the only notable change in $\langle v \rangle$ (solid blue line) between Figures \ref{fig:Fig2}(b)-(c)-(d) is the significant increase in its magnitude as the eye is approached, almost doubling in magnitude between Figures \ref{fig:Fig2}(d) and (b). The mean radial velocity -- $\langle u \rangle$ (solid red line) -- follows a similar trend, except that its magnitude is seen to reduce as towards the inner eyewall (Fig. \ref{fig:Fig2}(b)) compared to the outer eyewall (Fig. \ref{fig:Fig2}(c)). Similarly, the vertical velocity -- $\langle w \rangle$ (solid black line) multiplied by a factor of 10 -- remains negligible at $\approx$63m in the inflowing BL (Fig. \ref{fig:Fig2}(d)), increasing slightly in the outer eyewall (Fig. \ref{fig:Fig2}(c)), until its peak as the inner edge of the eyewall is approached (Fig. \ref{fig:Fig2}(b)). In summary, the most significant changes in the mean velocity fields between the inflowing BL and the eyewall (inner and outer), appears to be a mere amplification of the radial and tangential velocity fields, with the mean vertical velocity being comparatively negligible.
We note that at z$\geq$750m, the difference between the profiles in Figs. \ref{fig:Fig2}(b)-(c)-(d) are more complicated than a mere increase: Specifically, in the inner eyewall, $\langle u \rangle$ switches from inflow (negative) to outflow (positive) while $\langle w \rangle$ becomes non-negligible. We suspect that the findings detailed in the rest of this study (focused on the near surface region, z$\approx$63m) may therefore differ from greater heights where the mean conditions are much complicated. In this sense, the present study aims to provide a base understanding for future work.

Fig. \ref{fig:Fig2}(e) shows horizontal profiles of the azimuthally and time averaged vertical momentum fluxes (red line --  $\langle u^\prime w^\prime \rangle$, blue line --  $\langle v^\prime w^\prime \rangle$). It is clear that: in most of the near surface HBL, the mean vertical momentum flux remains mostly negative from the inflowing BL to the inner eyewall. A comparison of the red and blue lines indicate that: in most of the inflowing BL, the vertical momentum flux of tangential velocity ($\langle v^\prime w^\prime \rangle$) is comparatively larger than for the radial velocity ($\langle u^\prime w^\prime \rangle$), which is near zero. As the outer eyewall is approached, $\langle u^\prime w^\prime \rangle$ increases steadily until it becomes comparable to $\langle v^\prime w^\prime \rangle$ in the inner eyewall. These profiles do not, however, offer up any information on the kinematics of the flow peculiar to extreme cases of instantaneous fluxes which have been averaged out to produce these profiles, nor a breakdown of the negative fluxes (i.e., -$u^\prime w^\prime$ and -$v^\prime w^\prime$). To address clarify these further, we explore the quadrant distribution of the fluctuating velocity fields below.

\subsection{Quadrant analysis }\label{sec:sec3ddy}
\begin{figure}
    \centering
        \includegraphics[width=0.92\textwidth]{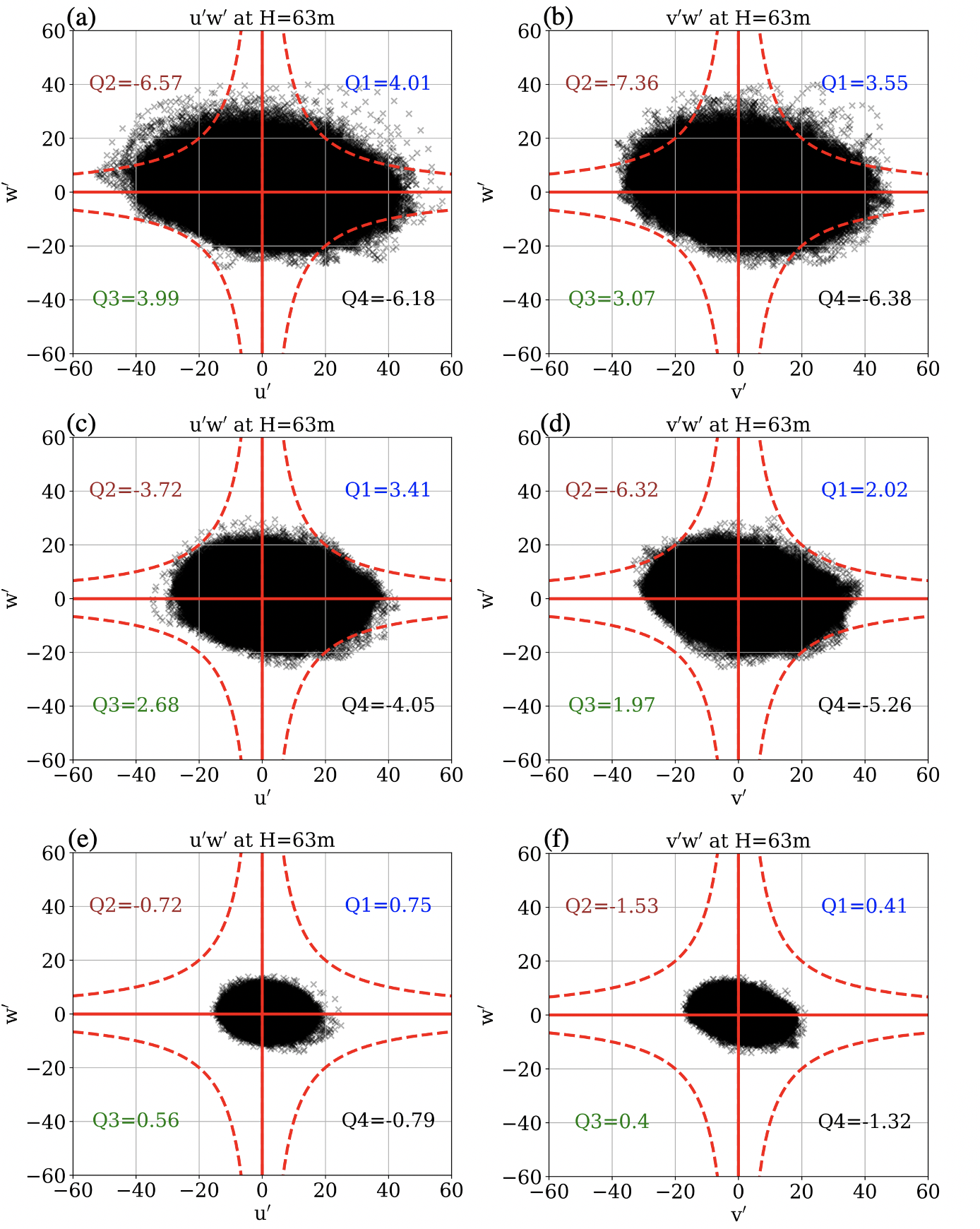}
    \caption{Quadrant distribution of fluctuating velocity components ($u^\prime$, $v^\prime$ and $w^\prime$) in the inner eyewall (r$\approx$10 - 13km -- (a),(b)),  outer eyewall (r$\approx$14 - 17km -- (c),(d)) and inflowing boundary layer (r$\approx$22 - 30km -- (e),(f)), at 63m height above the surface. The red dashed curves in each panel represents arbirary hyperbolic curves to visually identify extreme events. The blue, brown, green and black numbers in each quadrant denotes the mean flux components from the Q$_1$, Q$_2$, Q$_3$ and Q$_4$ quadrant respectively.}
    \label{fig:Fig3}
\end{figure}

Quadrant analysis, introduced first by \cite{wallace1972wall} and employed extensively in turbulence research \citep{shaw1983structure,robinson1989review,robinson1991kinematics,katul1997turbulent,nolan2010quadrant,wallace2016quadrant,li2024structure} is useful in addressing the issue above, in that the negative vertical momentum fluxes (i.e., -$\langle u^\prime w^\prime \rangle$ and -$\langle v^\prime w^\prime \rangle$) can be further broken down into ejection/sweep pairs (i.e., Q$_2$: [$-u^\prime$, $+w^\prime$], [ $-v^\prime$, $+w^\prime$] and Q$_4$: [$+u^\prime$, $-w^\prime$], [$+v^\prime$, $-w^\prime$] respectively) in addition to the so called "outward" (Q$_1$: [$+u^\prime$, $+w^\prime$], [ $+v^\prime$, $+w^\prime$]) and "inward" (Q$_3$: [$-u^\prime$, $-w^\prime$], [ $-v^\prime$, $-w^\prime$]) interactions. In the context of the mean fields discussed in Figs. \ref{fig:Fig2}(a)-(d), an ejection of radial velocity fluctuation (i.e., [$-u^\prime$, $+w^\prime$]) implies stronger-than-average inward flowing air parcels are being tossed upwards by positive fluctuating vertical velocity, while sweep events (i.e., [$+u^\prime$, $-w^\prime$]) indicate ouflowing air aloft being pulled down to the surface by negative fluctuating vertical velocity. Similarly, an ejection (sweep) of tangential velocity fluctuations indicate low (high) momentum air being lofted (pulled) upwards (downwards).
In this section we take the analysis presented in Fig. \ref{fig:Fig2}(e) a step further by investigating individual contributions of different quadrants to the mean vertical momentum flux. In all panels shown below (Fig. \ref{fig:Fig3}), the mean flux in each quadrant is noted in blue, brown, green and black, while the dashed red line is a hyperbolic function used to identify extreme instantaneous events.

Figures \ref{fig:Fig3}(a) and (b) show that the mean fluxes in the Q$_2$ and Q$_4$ quadrants dominate mean flux fields. A closer look at the scatter plot in both figures indicate that there are significant extreme Q$_1$ events i.e., [$+u^\prime$, $+w^\prime$] and [$+v^\prime$, $+w^\prime$] exceeding the dash red lines. This suggests that these few extreme cases, though visually noticeable, are still insufficient to account for a mean greater flux than seen for the Q$_2$/Q$_4$ quadrants. Using the same hyperbolic function in Figures \ref{fig:Fig3}(c) and (d), it is seen that there are less extreme events exceeding the dashed red lines in the outer eyewall than in the inner eyewall (Figures \ref{fig:Fig3}(a) and (b)). In addition, we note that Q$_2$/Q$_4$ quadrant events dominate the mean flux fields in the outer eyewall, similar to the inner eyewall. Similarly, in the inflowing BL, Figures \ref{fig:Fig3}(e) and (f) shows that there are comparatively much less extreme occurrences of extreme fluxes. In Fig. \ref{fig:Fig3}(e), the Q$_2$/Q$_4$ quadrant pair dominate, however, we also note that the mean flux from Q$_1$ events (top right quadrant in blue) slightly exceeds the ejection quadrant (Q$_2$) event. In Fig. \ref{fig:Fig3}(f), the Q$_2$/Q$_4$ events significantly dominate the mean flux contribution compared to other quadrants.

Results from the quadrant analyses above suggests that Q$_2$/Q$_4$ events possess the dominant flux contributions. The question thus arises as to what the structure of turbulent eddies associated with these events look like? In the following subsections, we focus on extreme ejection events as a criteria for conditional averaging. By examining the conditionally averaged perturbation fields around these events, the representative structure of the derived "conditional eddy" is examined. We then employ the well known Q-criteria \citep{hunt1988eddies,jeong1995identification,menon2021significance} for the identification of vortices, to further investigate the three dimensional structure of the conditional eddy. The Q-criteria identifies a vortex core simply as regions where the second invariant of the velocity gradient tensor, $$
Q=\left(\|{\boldsymbol{\Omega}}\|^2-\|{\boldsymbol{S}}\|^2\right) / 2
$$ is positive, where $\boldsymbol{\Omega}$ and $\boldsymbol{S}$ are the rotation and strain rate respectively.

\subsection{The structure of coherent eddies}\label{sec:sec3a_statiaory}
We begin our investigation of the structure of conditional eddies in the near-surface HBL by first examining the structure of coherent \textbf{ejection} events (i.e., [$-u^\prime$, $+w^\prime$] and [$-v^\prime$, $+w^\prime$]) in the inner eyewall (r$\approx$10 - 13km), outer eyewall (r$\approx$14 -- 17km) and inflowing BL (r$\approx$22 -- 30km). Using the conditional averaging scheme detailed in the previous section, we search for events in each of the 61 snapshots that satisfy the conditions: $-u^\prime$$+w^\prime$, $-v^\prime$$+w^\prime$ \textbf{and} $\sqrt{(u^{\prime}w^{\prime})^{2} + (v^{\prime}w^{\prime})^{2}} > 99.9\%$ i.e., ejection events (for radial and tangential velocity) whose mean vertical fluxes are in the top 99.9 percentile. 106, 99 and 271 events are identified in the inner eyewall, outer eyewall and inflowing BL respectively. These events are then averaged in each region. It should be noted that the resulting conditionally averaged fields are normalized by the local surface friction velocity ($u^{*}$ in m/s). A similar procedure is carried out for sweep events (see appendix D) with 107, 122 and 379 extreme events identified in the inner eyewall, outer eyewall and inflowing BL respectively.

\begin{figure}
    \centering
         \includegraphics[width=0.5\textwidth]{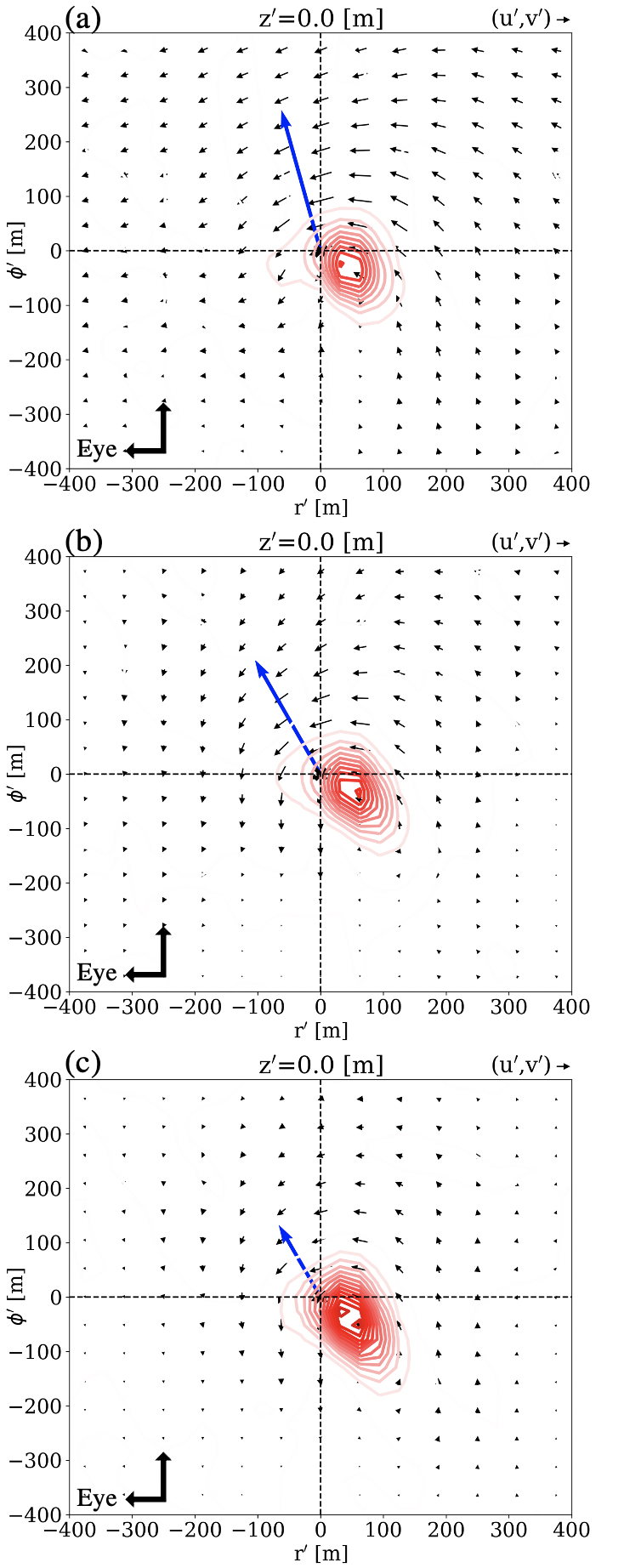}
    \caption{Horizontal sections through the center of the normalized conditionally-averaged structure (denoted by "x")  in the (a) inner eyewall, (b) outer eyewall and (c) inflowing BL. The vectors [($u^\prime, v^\prime$)], at the top right corner of each panel, represents the fluctuating velocity components (exaggerated by $\times$7 for visibility), while the red contours indicate the Q-criteria with [min, max, interval] = [-15,15,0.75]. The blue arrow in panel indicates the mean flow direction at the center of the averaging radii.}
    \label{fig:Fig4}
\end{figure}

Figures \ref{fig:Fig4}(a)-(c) shows a horizontal section through the conditionally averaged structure in the inner eyewall, outer eyewall and inflowing BL respectively. We note that the overall shape (as indicated by the red contour -- Q-criteria) of the conditional eddy in all three regions of the near-surface HBL are qualitatively similar (i.e., a vortex core inclined, in the direction of the tangential winds ($\phi$), towards the TC eye), indicating that the general structure of the conditional eddy associated with extreme vertical momentum flux does not depend significantly on the near-surface location in the HBL. We note, however, that the circulation (black vectors) around the vortex core increases from the inflowing BL (Fig. \ref{fig:Fig4}(c)) to the inner eyewall (Fig. \ref{fig:Fig4}(a)). The solid blue arrow at the center of the horizontal cross sections (Fig. \ref{fig:Fig4}(a), (b), (c)) points in the direction of the mean flow, with an inflow angle of -15.59$^{\circ}$, -29.85$^{\circ}$ and -30.07$^{\circ}$ respectively, while the length of the arrow indicates the mean velocity field. It is seen that the structures are roughly inclined with the mean flow. Seeing that the conditional eddies are qualitatively similar in all three regions analysed in this study, the rest of our discussion is focussed on the inner eyewall (Fig. \ref{fig:Fig4}(a)). A closer look at the wind vectors around the vortex core in Fig. \ref{fig:Fig4}(a) indicates that the characteristic conditional eddy associated with ejection events within the eyewall acts to "turn" the tangential flow ($\phi^{\prime}$ direction) of air parcels towards the TC eye ($r^{\prime}$ direction).

\begin{figure}
    \centering
        \includegraphics[width=1.0\textwidth]{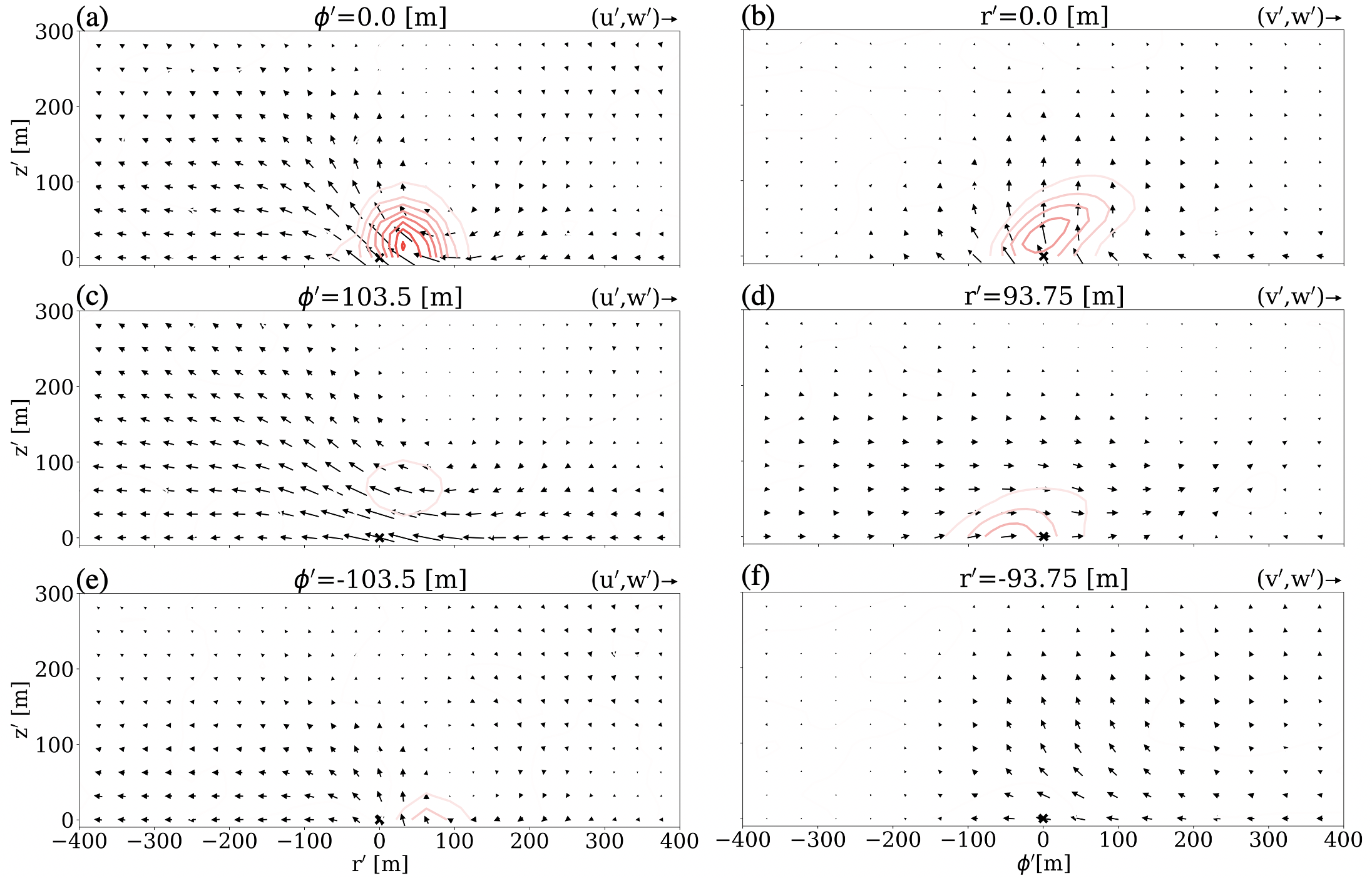}
    \caption{Transverse -- [(b), (d), (f)] and longitudinal -- [(c), (e), (g)]  slices through the center of the normalized conditionally-averaged structure (denoted by "x") in the inner eyewall (Fig. \ref{fig:Fig4}(a) above). The vectors [($u^\prime, w^\prime$), ($v^\prime, w^\prime$)] respectively, at the top right corner of each panel, represents the fluctuating velocity components (exaggerated by $\times$7 for visibility), while the red contours indicate the Q-criteria with [min, max, interval] = [-15,15,0.75].}
    \label{fig:Fig5}
\end{figure}

\begin{figure}
    \centering
        \includegraphics[width=0.58\textwidth]{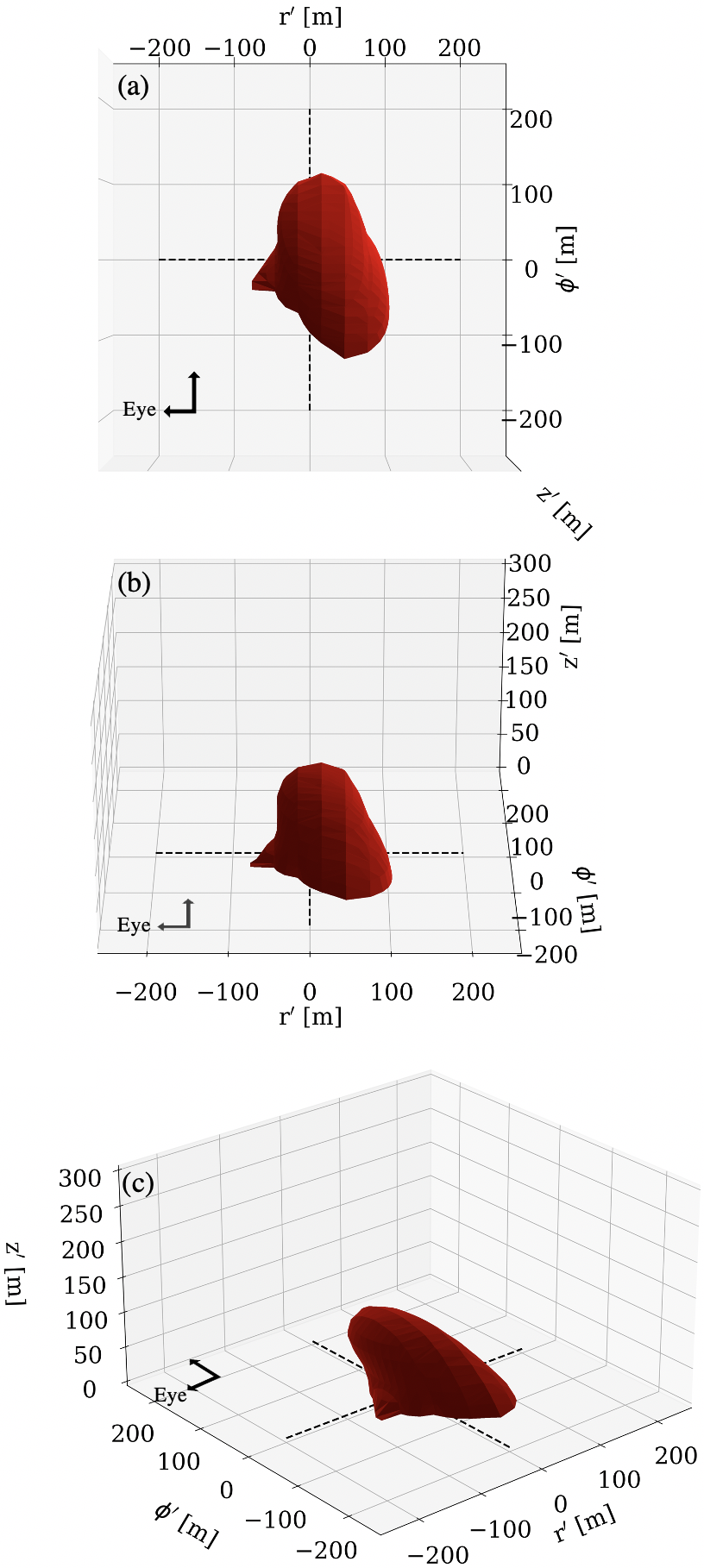}
    \caption{3D isosurfaces of Q-criteria=1.0 in top view -- (a) and side view -- (b) for the conditionally averaged structure in the inner eyewall, shown in Fig. \ref{fig:Fig4}(a). The dashed black lines cut across the centroid of the structure.}
    \label{fig:Fig6}
\end{figure}

To further understand the kinematics of the flow around the structure in Fig. \ref{fig:Fig4}(a), vertical cross sections through the center of the structure ($r^{\prime}=0$, $\phi^{\prime}=0$) as well as downstream (in-front) and upstream (behind) the center are shown in Fig. \ref{fig:Fig5}(a)-(f). From Fig. \ref{fig:Fig5}(a), we note that the structure associated with the "overturning" of inflowing air upwards in the inner eyewall is a single vortex core. This overturning of the inflowing air upwards is seen to decrease downstream of the structure ($\phi^{\prime}=103.5m$ -- Fig. \ref{fig:Fig5}(c)), occurring at an elevated height. This indicates that the head of the structure might be slightly lifted downstream. On the other hand, a vertical cross section upstream of conditional eddy structure ($\phi^{\prime}=-103.5m$ -- Fig. \ref{fig:Fig5}(e)) shows little overturning effect of the vortex core. A vertical cross section at $r^{\prime}=0$ (Fig. \ref{fig:Fig5}(b)) indicates that air parcels flowing in the tangential direction ($\phi^{\prime}$) tend to be lifted around the the vortex core. There is notably less overturning vortex signature in this direction (i.e., $\phi^{\prime}$ -- Fig. \ref{fig:Fig5}(b)) compared to the cross section at $r^{\prime}$ (Fig. \ref{fig:Fig5}(a)). A comparison of the vector fields downstream ($r^{\prime}=-93.5m$ -- Fig. \ref{fig:Fig5}(f)) and upstream ($r^{\prime}=93.5m$ -- Fig. \ref{fig:Fig5}(d)) shows opposing flow close to the conditional eddy, confirming a vertical vorticity component to the rotation seen in Fig. \ref{fig:Fig4}(a).

The foregoing discussion suggests the presence of a three dimensional vortex, lofting inflowing air upwards and circulating horizontal flow about it's core. In order to clarify the nature of this conditional eddy (shown in Figs. \ref{fig:Fig4}(a) and \ref{fig:Fig5}(a)-(f)), several views of the 3D isosurface of Q-criteria=1.0 is shown in  Fig. \ref{fig:Fig6}. The plan view (Fig. \ref{fig:Fig6}(a)) of the conditionally averaged structure clearly confirms the horizontal inclination of the vortex core towards the TC eye, as seen in the two-dimensional horizontal slice (Fig. \ref{fig:Fig4}(a)). The oblique views in Figs. \ref{fig:Fig6}(b) and (c) confirms the vertical extent of the vortex core to be approximately 100m in height, confirming the two-dimensional cross section in Fig. \ref{fig:Fig5}(a). Most interestingly, Fig. \ref{fig:Fig6}(c) shows very clearly that the "head" of the coherent vortex core is lifted downstream, further clarifying the vertical cross sections in Figs. \ref{fig:Fig5}(a) and (c) where the "over turning" influence of the vortex core is lifted upwards downstream (i.e., $\phi^{\prime}=103.5m$). Overall, we find (from Figs. \ref{fig:Fig4}, \ref{fig:Fig5} and Fig. \ref{fig:Fig6}) that the characteristic eddy associated with near-surface extreme vertical momentum fluxes (for ejection events) is a single vortex core, roughly horizontally inclined in the direction of the mean winds. We note that eddy conditioned on sweep events (see appendix D) exhibit an opposing flow (compared to ejections) around the structure. However, the conditional eddy appears less coherent that the Q-criteria signatures are weak.

\section{Discussion and Conclusions}\label{sec:sect4_conclusions}
In this study, output from a LES ($\Delta x = \Delta y  $ = 31.25m, \  $\Delta z$ = 15.625m) of an idealized Category 5 TC is analyzed to investigate near-surface behaviour of coherent structures in the HBL and their relationship to extreme vertical momentum flux occurrences. In order to overcome the certain limitations of "bulk" approaches commonly used in the analysis of turbulent fluxes in the meteorology research community -- e.g. the spatial and temporal averaging over the entire HBL as in \cite{oguejiofor2024role} -- conditional averaging was employed to educe the spatial structure of eddies associated with the most extreme flux occurrence. Using quadrant analysis, we confirm first the dominance of ejection/sweep events and then we conditionally average the spatial field around ejection events that are collocated with extreme vertical momentum fluxes. We find that the conditionally averaged structure exists as a vortex, roughly inclined with the mean winds. The structure associated with extreme sweep events appear to be comparatively less spatially coherent. The kinematics of the vortex core found in this study agrees with the mechanism (i.e., fluid acceleration) suggested by \cite{bernard1990reynolds}, \cite{handler1992role} and \cite{bernard1993vortex} for the generation of extreme Reynolds stress events. 

The coherent structure educed in this study differs from the "double-roller" signature found in LESs of atmospheric boundary layer flows \citep{lin1996coherent, foster2006near} and hairpin vortices in vegetation canopies \citep{finnigan2009turbulence,li2024structure}. In their LES of a neutrally stratified PBL, \cite{foster2006near} rotated both vertical momentum flux components (i.e., $v^{\prime}w^{\prime}$ and $u^{\prime}w^{\prime}$) to be aligned. We choose, instead, to investigate the mean flux instead ($\sqrt{(u^{\prime}w^{\prime})^{2} + (v^{\prime}w^{\prime})^{2}}$) and believe that this subtle difference is not responsible for the difference in the structures found. We also explored (not shown in this study) structures conditioned on extreme vertical momentum fluxes alone and separately (i.e., $-v^{\prime},+w^{\prime}<0$ and $-u^{\prime},+w^{\prime}<0$) and found a pair of counter-rotating legs of inclined vortices in the former and a single vortex core in the later. We insist that the interpretation of the vertical momentum flux of tangential velocity ($-v^{\prime},+w^{\prime}$) separately from radial velocity ($-u^{\prime},+w^{\prime}$) is not justified seeing that some combination of both quantities (in our case -- $\sqrt{(-u^{\prime},+w^{\prime})^{2} + (-v^{\prime},+w^{\prime})^{2}}$) defines the mean vertical momentum flux. Furthermore, we confirm the robustness of the structure educed in this study by a series of sensitivity experiments varying the threshold for identifying extreme mean vertical momentum flux (i.e., the 99.9 percentile shown in this study).

Although this study focused on near-surface ejection events for the vertical momentum fluxes, we have shown key evidence on the a coherent vortex core being the characteristic eddy structure in the inner eyewall, outer eyewall in inflowing BL. From Figs. \ref{fig:Fig2}(a)-(d), we re-emphasize that the height of focus in the present study is the near-surface (h=63m) slightly above the height of maximum radial inflow, where the mean vertical velocity is comparatively minimal and the radial variation in radial and tangential velocity is simpler. This is not the case at z$\geq$750m. Future research efforts thus lie in the investigation of the kinematics of coherent structures associated with extreme vertical momentum fluxes far from the near-surface region of the HBL, possibly in the much more complicated inflow-outflow zone within the eyewall. More research effort could also be directed towards the investigation of near-surface coherent structures in intense TCs with moderate to strong vertical wind shear (unlike the present study with where the storm is symmetric).

\section*{Acknowledgments}
The authors would like to acknowledge useful personal correspondence with Dr. Richard Rotunno, Peter P. Sullivan, Edward (Ned) Patton, Robert H. Shaw and John Finnigan. We also acknowledge ONR grants N00014-19-S-B001 to the University of Notre Dame and N00014-20-1-2071 to NCAR/UCAR for financial support under the Tropical Cyclone Rapid Intensification (TCRI) campaign.
George Bryan was supported by the NSF under Cooperative Agreement No. 1852977. 
We also acknowledge high-performance computing support from Cheyenne (doi:10.5065/D6RX99HX) provided by NSF NCAR's Computational and Information Systems Laboratory.

Due to the large size (several tens of terabytes), the model simulation output used in this study is stored on NSF NCAR's campaign storage system, and is available upon request to Dr. George Bryan.

\appendix
\section{Structure of conditionally averaged sweep events}\label{appA}
\begin{figure}
    \centering
         \includegraphics[width=1.0\textwidth]{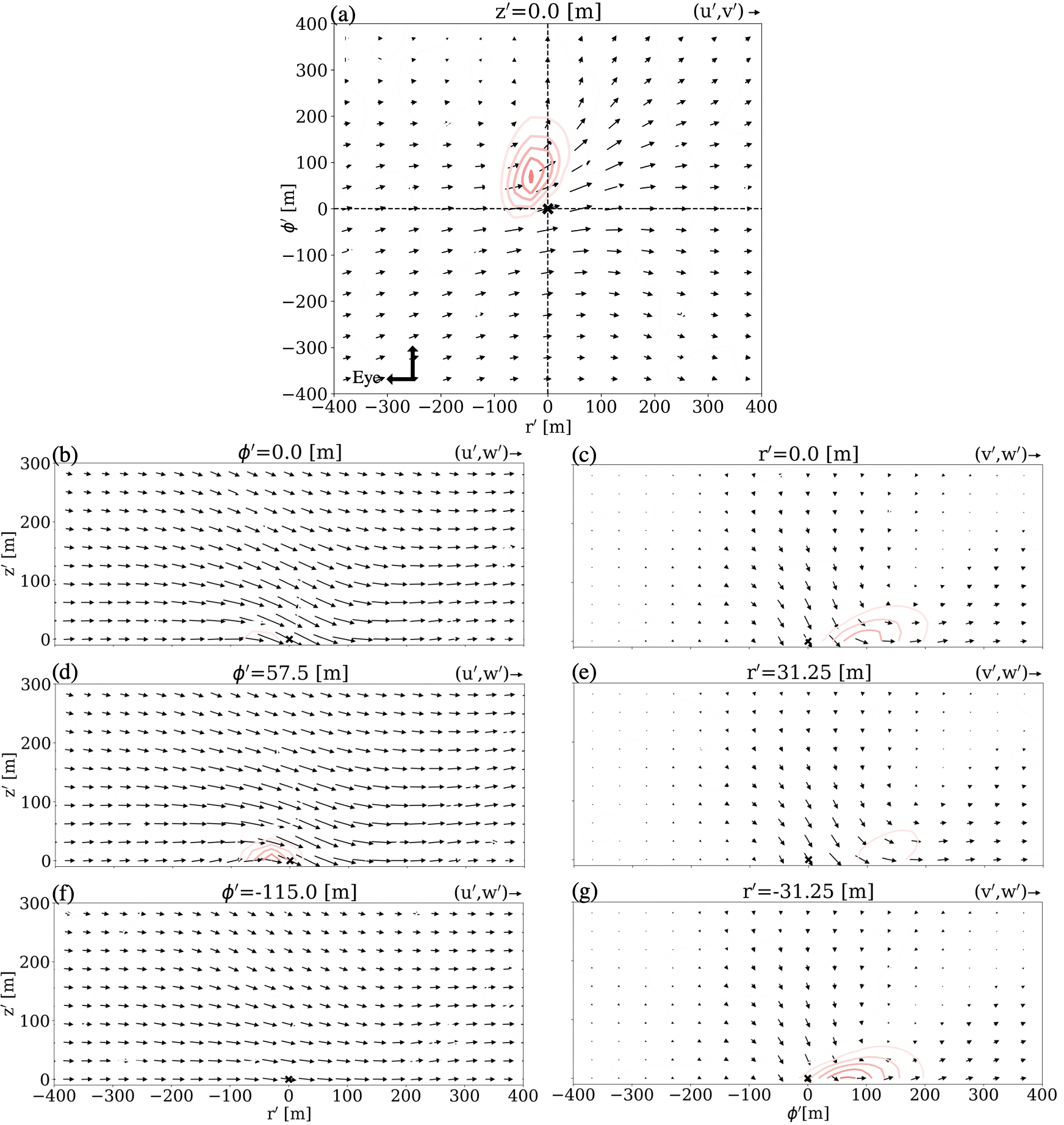}
    \caption{Same as Fig. \ref{fig:Fig4}(a) and Fig. \ref{fig:Fig5}, but for sweep events i.e., Q$_4$: [$+u^\prime$, $-w^\prime$], [$+v^\prime$, $-w^\prime$].}
    \label{fig:Fig_appA}
\end{figure}

Figures \ref{fig:Fig_appA} shows horizontal (a) and vertical ((b)-(g)) cross sections through the "eddy" conditioned on the occurrence of extreme sweep (Q$_4$) events, similar to Figs. \ref{fig:Fig4}(a) and Fig. \ref{fig:Fig5}(a)-(f) respectively. Fig \ref{fig:Fig_appA}(a) shows that there is less rotation (as seen in the flow vectors) around the structure as flow moves away from the TC eye (compared to Fig. \ref{fig:Fig4}(a)). Vertical cross sections in the $\phi^{\prime}$ direction (Figs. \ref{fig:Fig_appA}(b), (d) and (f)) shows flow vectors away from the TC eye (from  $- r^{\prime}$ to  $+r^{\prime}$) being pulled down around the center. Furthermore,vertical cross sections in the $r^{\prime}$ direction (Figs. \ref{fig:Fig_appA}(c), (e) and (g)) shows the action of the coherent vortex in pulling down flow from  $- \phi^{\prime}$ to  $+\phi^{\prime}$.

\clearpage
\bibliographystyle{jfm}
\bibliography{jfm-instructions}

\end{document}